\documentclass[prb,citeautoscript,superscriptaddress,twocolumn]{revtex4}

\usepackage{amsfonts,amsmath,amssymb}
\usepackage[dvips]{graphicx,color}
\input epsf 

\newcommand{\dir}{Figs}

\begin{document}
\title{Thermal Fluctuations in a Lamellar Phase of a
Binary Amphiphile-Solvent Mixture: \\ a  Molecular Dynamics Study.} 

\author{C. Loison}
\email{claire@cecam.fr}
\affiliation{Fakult\"at f\"ur Physik, Universit\"at Bielefeld,
Universit\"atsstra{\ss}e 25, D-33615 Bielefeld, Germany}
\affiliation{Centre Europ\'een de Calcul Atomique et Mol\'eculaire,
ENS Lyon, 46, All\'ee d'Italie, 69007 Lyon, France}

\author{M. Mareschal}
\affiliation{Centre Europ\'een de Calcul Atomique et Mol\'eculaire,
ENS Lyon, 46, All\'ee d'Italie, 69007 Lyon, France}

\author{K. Kremer}
\affiliation{Max Planck Institute for Polymer Research,
Ackermannweg 10, 55128 Mainz, Germany}

\author{F. Schmid}
\email{schmid@physik.uni-bielefeld.de}
\affiliation{Fakult\"at f\"ur Physik, Universit\"at Bielefeld,
Universit\"atsstra{\ss}e 25, D-33615 Bielefeld, Germany}

\date{version of \today}
\begin{abstract}
\noindent
We investigate thermal fluctuations in a smectic A phase of an
amphiphile-solvent mixture with molecular dynamics simulations. 
We use an idealized model system, where solvent particles are represented 
by simple beads, and amphiphiles by bead-and-spring tetramers. 
At a solvent bead fraction of 20 \% and sufficiently low temperature, 
the amphiphiles self-assemble into a highly oriented lamellar phase.  
Our study aims at comparing the structure of this phase with the 
predictions of the elastic theory of thermally fluctuating fluid 
membrane stacks [Lei {\it et al.}, J. Phys. II {\bf 5},1155 (1995)]. 
We suggest a method which permits to calculate the bending rigidity 
and compressibility modulus of the lamellar stack from the simulation 
data. The simulation results are in reasonable agreement with the theory.
\end{abstract}

\maketitle


\section{Introduction. }

Lipids are essential components of biomembranes. Their ability to self-assemble 
into bilayers is characteristic for amphiphilic molecules, {\em i.e.}, molecules 
with a hydrophilic head-group and one or several hydrophobic tails. 
In concentrated aqueous solution, most lipids form a lamellar $L_\alpha$ 
phase: a stack of amphiphile bilayers separated by layers of solvent. At room 
temperature, the bilayers usually have the structure of two dimensional fluids. 
The bilayer stack 
exhibits liquid-like behavior in two directions, and (quasi)-crystalline 
ordering in the direction perpendicular to the layers. Therefore, the $L_\alpha$ 
lamellar phase can be described as a smectic liquid crystal. 
The bilayers are planar on average, with a well defined inter-layer spacing which 
can be measured by X-Ray diffraction. In addition to this positional ordering, 
the molecules exhibit orientational ordering perpendicular to the lamellar
plane (smectic A).

From an experimental point of view, lamellar phases are useful model
systems which allow to study the structure of lipid bilayers very
conveniently, {\it e.g.} in diffraction studies. The shape of X-ray diffraction 
peaks has been discussed mostly in terms of the classical theory of smectic A, 
as developed originally by  Caill\'e \cite{Caille_AcS_72} and de Gennes 
\cite{deGennes_93}, and further elaborated by Lei {\it et al.} \cite{Lei_JP_95}. 
This is a continuum approach, which operates on the mesoscopic level and 
describes the lamellar material as a stack of two-dimensional fluctuating layers.
The free energy is taken to be an elastic energy, which penalizes local 
layer deformations and local deviations from the average interlayer distance. 
Theories of this type have been used to measure the bending constant $K$ and 
the compressibility $B$ in smectics. Applied to highly aligned experimental
samples, they even allowed to calculate the bending rigidity of single bilayers, 
and the effective interactions between them \cite{Bouglet_EPJB_99}.

On molecular length scales, interfaces in complex fluids 
can also be investigated by molecular simulations 
\cite{Mueller_JCP_96,Goetz_PRL_99,Werner_PRE_99,Werner_JCP_99,Marrink_JCB_01}.
The simulation results can then be used to verify the validity of 
mesoscopic
theories. For example, the phenomenological description of single bilayers
in terms of a surface tension $\gamma$ and a bending rigidity $K_c$ has been
tested for idealized amphiphile models \cite{Goetz_PRL_99}, and more recently
even for a realistic phospholipid model \cite{Marrink_JCB_01}.
The present work aims at extending this type of study to entire lamellar
stacks of bilayers. To this end, we have performed large scale molecular 
dynamics 
simulations of a simplified coarse-grained model for binary amphiphile-solvent 
mixtures. This made it possible to study stacks of up to fifteen
bilayers, systems large enough to be compared to the continuum theory
for smectics mentioned above.
A straightforward analysis,
based on the direct inspection of the structure factor, failed because 
it requires data with a very small statistical error. We have developed an 
alternative, more robust method, which allowed us to extract the 
phenomenological parameters $K$ and $B$. On large length scales, 
our simulation results agree well with the theory.

The paper is organized as follows: In the next section, we recall the principal
features of the theory (the ``discrete harmonic model'',
\cite{Lei_JP_95,Holyst_PRA_91}).  In section \ref{model}, we introduce the 
simulation model and describe the simulation method, and section \ref{results} 
contains our results. There we first discuss briefly the phase behavior
of our model (\ref{results_phase}). 
Then we analyze the bilayer fluctuations in a lamellar phase. The fluctuations 
about the mean position of each membrane give the bending energy of the 
bilayers, and the correlations of fluctuations between adjacent membranes 
yield the interactions between membranes. 
We summarize and conclude in section \ref{conclusions}.

\section{Theoretical background: Elasticity in smectic A.}
\label{theory}

Before discussing the simulations, we briefly sketch the continuum theory
which we use to analyze the data. It describes the system by a discrete 
set of layers, stacked in the $z$-direction and extending continuously in the 
$(x,y)$-plane. The average distance between layers is $\bar{d}$. We assume 
that we can parametrize each layer $n$ by a unique height function 
$z=h_n(x,y)$, and that the molecules in a layer are perpendicular 
to the surface, {\it i.e.}, the local director is given by the layer normal. 
The fluctuations about the mean position of the layer are characterized by 
the local displacement $u_n(x,y) = h_n(x,y) - n \bar{d}$.

Most generally, deformations of smectics may include twist, bend, and splay 
modes of the director, and compression of the layers. However, the energy 
penalty on twist and bend modes is very high, because these modes cannot be 
realized at constant layer spacing. Therefore, they are effectively suppressed, 
and the remaining relevant deformations are the splay mode and the layer 
compression \cite{deGennes_93,Chaikin_95,Chandrasekhar_77}. 
The problem can be further simplified
by adopting the "discrete harmonic" approximation (DH), which has been used
successfully to interpret X-ray scattering data of highly oriented 
lamellar phases \cite{Lei_JP_95} and to study the interfacial properties
of thin films of the lamellar phase \cite{Holyst_PRA_91}. Here only interactions
between adjacent layers are taken into consideration, and the free
energy is approximated by
\begin{eqnarray}
\label{Fd}
{\cal F}_{DH} &= &  \sum_{n=0}^{N-1}  \int_A \text{d}x \text{d}y
\left\{
\frac{K_c}{2} \left( \frac{\partial^2 u_n}{\partial x ^2} 
+\frac{\partial^2 u_n}{\partial y ^2}  \right)^2 \right.
\nonumber \\ &&
\left.
+ \:
 \frac{B}{2} \left(  u_n -u_{n+1} \right)^2 
\right\}.
\end{eqnarray}
The first term accounts for the bending energy of individual bilayers, 
and the second term approximates the free energy of compression.
(We note that layer bending should not be confused with the bend mode of 
the smectic, which is neglected here.)
The elasticity of the smectic phase is thus characterized by the two coefficients 
$K_c$, the bending modulus of a single membrane, and $B$, the compressibility
modulus. These are connected with the bulk compression modulus $\bar{B}$ and the 
bulk bending modulus $K$ by the simple relations $B = \bar{B}/\bar{d}$ and 
$K_c = K\bar{d}$. They define the in-plane correlation length 
$\xi = (K_c/B)^{1/4}$ and the characteristic smectic length 
$(K/\bar{B})^{1/2}$. The surface tension $\gamma$ of the
bilayers is taken to vanish, as in a  bulk phase.

The fluctuations of $u_n$ are most conveniently studied in Fourier space, 
because the Fourier modes decouple in (\ref{Fd}) and the equipartition 
theorem applies. We perform continuous Fourier transformations in the 
$x$- and $y$-direction, and a discrete Fourier transformation in the 
$z$-direction. This gives
\begin{eqnarray}
\label{TF1}
u(q_z,{\bf q}_\perp) & = &   \sum_n  u_n({\bf q}_\perp) ~e^{- i q_z n\bar{d} } \\
\label{TF2}
u_n({\bf q}_\perp) &=& \int_A  \text{d}r~ u_n({\bf r})~e^{- i {\bf q}_\perp {\bf r}}\\
\label{TF3}
&=& \frac{1}{N} \sum_{q_z} u(q_z,{\bf q}_\perp) ~e^{+ i q_z n\bar{d}},
\end{eqnarray}
where $q_z$ is the $z$-component of ${\bf q}$ and ${\bf q}_\perp$ the projection 
into the $(x,y)$-plane. In simulations, systems have finite extensions 
$L_x, L_y, L_z$ and periodic boundary conditions apply. The components of the 
${\bf q}$-vector then take only discrete values $q_\alpha = k_\alpha (2\pi)/L_\alpha$ 
with integer $k_\alpha$. The maximum number of independent $z$-components $k_z$ 
is given by the number of bilayers $N$.

From the equipartition theorem, one can then calculate the average amplitudes 
of fluctuations for large systems \cite{Safran_94}
\begin{equation}
\label{Uq2}
\langle |u({\bf q}_\perp,q_z)|^2 \rangle
 = 
\frac{N L_x L_y ~k_BT}{ 2B \left[1-\cos(q_z\bar{d})\right] + K_c ~ q_\perp^4 }.
\end{equation}
Here and throughout, brackets $\langle \cdot \rangle$ refer to thermal averages. 
Unfortunately, the statistical error of our simulation results for
$\langle  |u({\bf q}_\perp,q_z)|^2 \rangle$ was too large to
allow for a direct comparison with Eq. (\ref{Uq2}). Therefore
we resorted to studying the integrated quantities

\begin{eqnarray}
s_n({\bf q_\perp}) &\doteq& 
\label{sum_coupled_spectra}
\frac{1}{N^2} \sum_{q_z} e^{iq_zn\bar{d}}~ \langle|u({\bf q_\perp},q_z)|^2 \rangle \\
\label{sum_coupled_spectra1}
&=&
\frac{1}{N}\sum_{j=0}^{N-1}~ \left \langle u_j({\bf q_\perp}). 
u_{n+j}({\bf q_\perp})^{\dag} \right\rangle. 
\end{eqnarray}
The quantity $s_0({\bf q_\perp})$ describes correlations within membranes, whereas 
$s_n({\bf q_\perp})$ (at $n > 0$) characterizes correlations between membranes.
In an infinitely thick stack of $N \to \infty$ bilayers, the sum $\sum_{q_z}$ 
can be replaced by the integral \mbox{$(N \bar{d}/2 \pi) \int_0^{2 \pi/\bar{d}} \text{d}q_z$}.
Inserting Eq. (\ref{Uq2}), we obtain
\begin{eqnarray}
\label{intUq1}
 s_0(q_\perp) & \stackrel{N \to \infty}{=}  &
\frac{L_x L_y~ k_BT}{K_c q_\perp^4}~\left[1 +\frac{4}{(\xi q_\perp)^4} \right]^{-1/2},\\
\label{sum_coupled_spectra2}
 s_n(q_\perp) & \stackrel{N \to \infty}{=}  &
 s_0(q_\perp) \times \left[ 1 + \frac{X}{2} - \frac{1}{2} \sqrt{X(X + 4)} \right]^n
\end{eqnarray}
where $\xi$ is the in-plane correlation length, $\xi = (K_c/B)^{1/4}$, and the
ratios $s_n/s_0$ between cross-correlations $s_n$ of membranes and the 
autocorrelation $s_0$ depend only on the dimensionless parameter 
$X = (\xi q_\perp)^4$. In deriving Eqs. (\ref{intUq1}) and 
(\ref{sum_coupled_spectra2}), we have made
use of the formula
\begin{equation}
\label{integral}
\frac{1}{2 \pi} \int_0^{2 \pi} d\tau \: \frac{e^{i n \tau}}{a - \cos(\tau)} 
= \frac{(a - \sqrt{a^2-1})^{n}}{\sqrt{a^2-1}},
\end{equation}
which can be derived by substituting $z = e^{i \tau}$ and applying the
residuum theorem.

Two regimes are expected with a crossover at $q_c \sim \xi^{-1}$. 
If $q_\perp$ is much larger than $q_c$, the fluctuation spectrum $s_0$ of
single membranes is proportional to $q_\perp^{-4}$. This corresponds 
to the well-known spectrum of single isolated membranes without 
surface tension. The relative amplitudes $s_n/s_0$ of cross-correlations
between different layers decay exponentially like $(1/X)^n$ with
the distance $n \bar{d}$ between the bilayers. In the small-wavelength regime,
fluctuations of different membranes are thus basically incoherent,
and the bilayers behave like free, unconstrained membranes.

In contrast, the regime $q_\perp \ll q_c$ is dominated by the coupling 
of compression modes with the bending fluctuations. The ratios $s_n/s_0$
tend towards one in the infinite wavelength limit $X \to 0$, {\it i.e.},
fluctuations of different bilayers are coherent. The fluctuation spectrum
is proportional to $q_\perp^{-2}$. The fluctuations thus grow more slowly
with the wavelength than those of free membranes, due to the fact that 
the membranes are constrained in the stack.

These results allow one to derive the height-height correlation
function, which has often been used to discuss X-ray scattering spectra
\cite{Lei_JP_95,Petrache_PRE_98,Lyatskaya_PRE_01}
$\langle \delta u_n(r)^2 \rangle$ with
\begin{equation}
\label{delta_unr_def}
\delta u_n(x,y)^2  \doteq	\frac{1}{N}\sum_{j=0}^{N-1}  |u_{n+j}(x,y) -u_j(0,0)|^2.
\end{equation}
This quantity can be calculated from back transforming
the $s_n(q_\perp)$ into the real space $(x,y)$. One obtains
\begin{equation}
\label{delta_unr}
\langle \delta u_n(r)^2 \rangle =
\frac{2 \eta_1}{q_1^2}
 \int_0^\infty \text{d}\tau~
\frac{1- J_0( \frac{r}{\xi} \sqrt{2\tau}) \left[\sqrt{1+\tau^2}-\tau\right]^{2n} }
{\tau\sqrt{1+\tau^2}},
\end{equation}
where $r = \sqrt{x^2 + y^2}$, $J_0$ is the first Bessel function,
$q_1$ the position of the first diffraction peak ($q_1 = 2 \pi/\bar{d}$),
and $\eta_1$ is the Caill\'e parameter \cite{Caille_AcS_72}
\begin{eqnarray}
\label{Caille}
\eta_1 & = &\frac{k_BT}{8 \sqrt{BK_c}}~\frac{4\pi}{\bar{d}^2}.
\end{eqnarray} 
The Caill\'e parameter is often used to characterize the width of
diffraction peaks. Within the elastic theory (\ref{Fd}), it
can be determined  easily, since all height-height correlations 
are proportional to $\eta_1$. 

\section{Model and method.}

\label{model}

\subsection{The simulation model.}

Our simulation model is based on a coarse-grained off-lattice model, 
which has been used extensively for polymers 
\cite{Grest_PRA_86,Kremer_PRL_88,Kremer_JCP_90}, and was recently 
optimized to study rheologic properties of amphiphilic dimers 
\cite{Soddemann_EPJE_01,Guo_PRE_02}. 

All molecules are represented by one or 
several soft beads (for simplicity, all beads are taken
 to have the same size $\sigma$ and the same mass $m$).
The solvent is represented by single soft spheres (type $s$). 
One solvent bead represents roughly three water molecules 
(the simulations could also be compared to amphiphiles in an 
oily solvent. In that case, one bead would correspond to one propane molecule, or to a
portion of a bigger alkane).
 The amphiphilic molecules are linear tetramers composed of
 two solvophobic beads (or "tail beads", denoted $t$) and
 two solvophilic beads (or "head beads", denoted $h$).  
The soft spheres of the amphiphilic $h_2t_2$ are covalently bonded.

\begin{figure}[h!]
\begin{center}
\includegraphics[width=8cm]{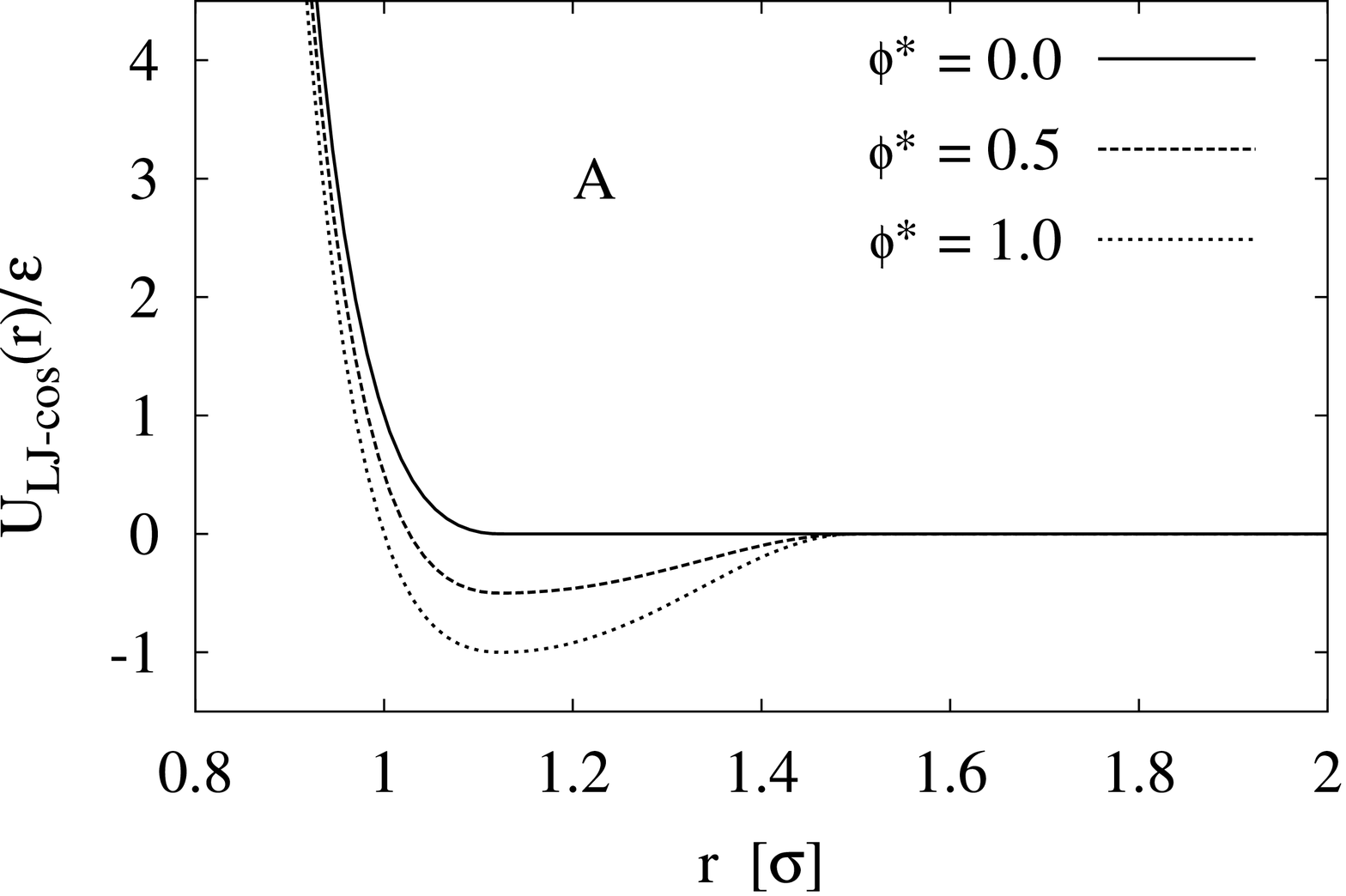}
\includegraphics[width=5.2cm,angle =-90]{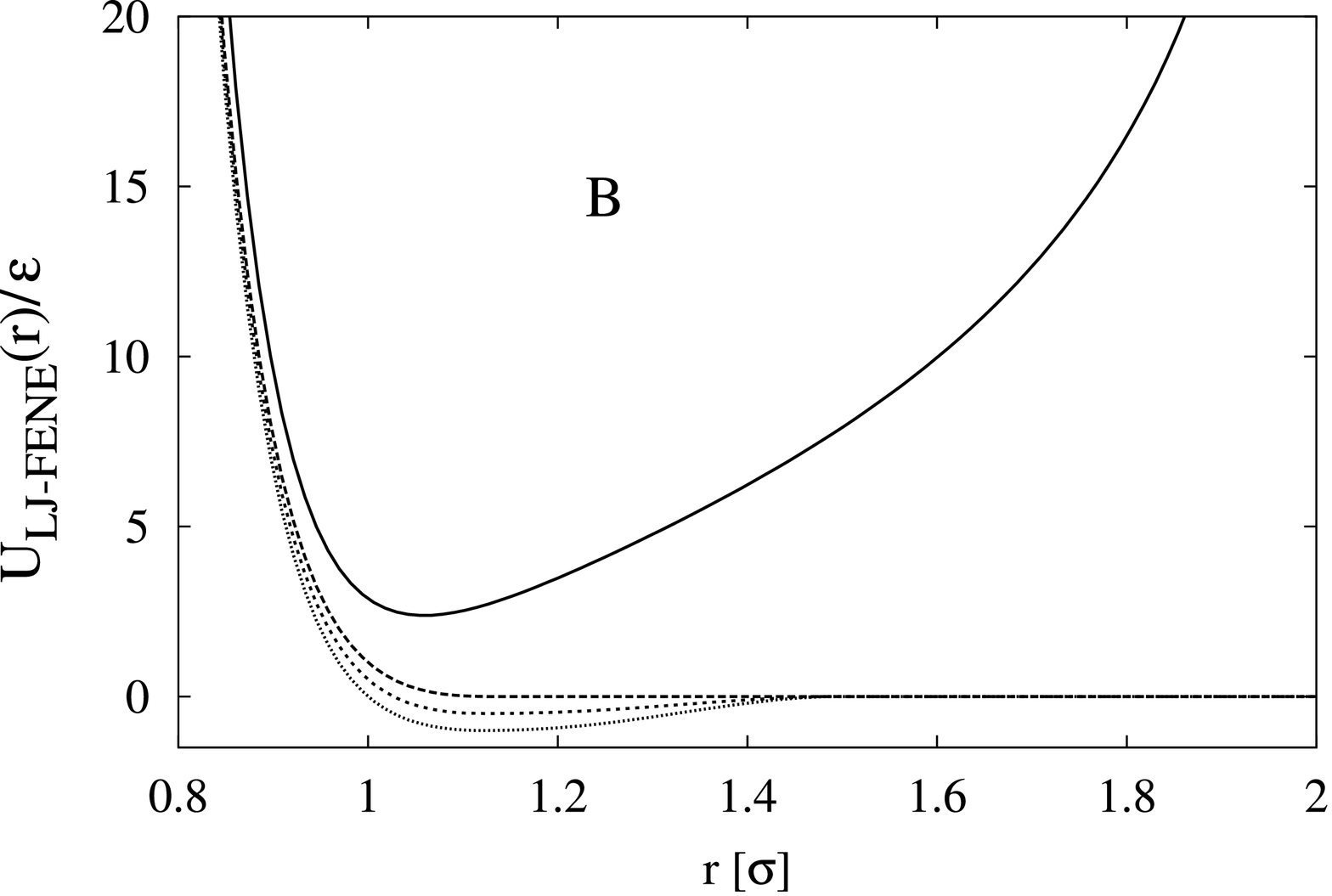}
\end{center}
\caption{Pair Potentials used in the simulations 
as a function of the
inter-particle distance in units of $\sigma$. 
{\bf A)}: Non-bonded interactions $U_{LJ-cos}$ for three choices of the 
potential depth $\phi$ ($\phi^* =\phi/\epsilon$). By construction, the position of the minimum 
($r = 2^{1/6} \,\sigma$) and the cut-off ($r_c =1.5 \,\sigma$) are independent 
of the potential depth. 
{\bf B)}: Bonded interactions between connected beads $U_{LJ-FENE}$ (solid line). 
The minimum is located at $r=1.06 \,\sigma$.
Dashed and dotted line show the non-bonded potentials for comparison.
}
\label{potentials}
\end{figure}

 Non-bonded beads interact with short ranged potentials of the form 
(see Fig. \ref{potentials} A for an illustration) 
 \begin{eqnarray}
\label{LJ-cos}
\lefteqn{ U_{LJ-cos}(r)=} & \\
&
\left\{
  \begin{array}{ll}
       4 \epsilon \left[\left(\frac{\sigma}{r}\right)^{12} 
       - \left(\frac{\sigma}{r}\right)^6 +\frac{1}{4}\right] 
       - \phi & \mbox{if } r  \leq 2^{1/6} \sigma\\ 
       \frac{\phi }{2}\left[ \cos(\alpha r^2+\beta) -1\right] 
       & \mbox{if } 2^{1/6} \sigma \leq  r\leq r_c \\
       0 & \mbox{if } r_c \leq r
     \end{array}
   \right. . \nonumber
\end{eqnarray}
where $\sigma$ is our unit of length and $\epsilon$ our unit of energy.
The potentials comprise a Lennard-Jones type soft repulsive part,
and a short-ranged attractive part. The parameters $\alpha$ and $\beta$ 
are fixed such that potentials and forces are continuous everywhere.
(\mbox{$\alpha = \pi/r_c^2-2^{1/3} \sigma^2$} and \mbox{$\beta=2 \pi -r_c^2 \alpha$}).
The energetic parameter $\phi$ determines the depth of the potential and the energetic parameter $\epsilon$ dertermines the strength of the soft 
repulsive core. At $\phi =0$, the interaction is purely repulsive.
The potential depth $\phi$ of the
pair interactions is the same for all pairs of beads which
``like'' each other ($ss, sh, hh$, and $tt$), and zero for pairs which 
``dislike'' each other ($ts$ and $th$). For a fixed $\epsilon$, the self-assembly is driven 
by the single energetic parameter $\phi$.  Unless stated otherwise, 
the parameter value for the pairs $ss, sh, hh$, and $tt$ is $\phi = 1.1 \,\epsilon$.
The range $r_c$ of the potential is chosen $r_c = 1.5\, \sigma$.

Bonded beads are connected by springs with the spring potential (see fig. \ref{potentials} B)
\begin{eqnarray}
\label{FENE_pot}
\lefteqn{U_{LJ-FENE}(r) =}&\\
&
\left\{
  \begin{array}{ll}
        4  \epsilon \left[\left(\frac{\sigma}{r}\right)^{12} 
    - \left(\frac{\sigma}{r}\right)^6\right] 
    - \left( \frac{\kappa  r_b^2}{2\sigma^2}\right)
    \ln\left[ 1-\left(\frac{r}{r_b}\right)^2\right]& 
    \mbox{if } r  \leq  r_b \\ 
    \infty & \mbox{if } r_b  \leq r \\
     \end{array}
   \right.
\nonumber
\end{eqnarray}
 named finite extendable nonlinear spring potential (FENE).
The bond parameters were fixed at $r_b = 2.0 \, \sigma$ and 
$\kappa = 7.0 \, \epsilon $. 

No chemical 
details are incorporated in the model. In particular, it contains no
long-range interactions, and no chain stiffness.

In the following, lengths shall be given in units of $\sigma$, 
energies in units of  $\epsilon$ and masses in  units of $m$. 
This gives the time unit $\tau =  (m \sigma^2 /\epsilon)^{1/2}$. 
Typical orders of magnitude of our units are 
$\epsilon \sim  5\cdot 10^{-21}\,$J, $m \sim 10^{-25}$ kg, $\sigma \sim 5 \AA$, and 
$\tau \sim 10^{-12}\,s$.

As has been discussed by Soddemann  {\it et al.}\cite{Soddemann_EPJE_01},
this coarse-grained model is simple enough that it permits to simulate very
large systems. In the present work, a smectic phase composed of up to fifteen 
bilayers, containing several thousands of molecules each, was simulated over 
about $10^5 \,\tau$ (about $100$ ns). 
Previous coarse-grained or all-atom simulations of bilayers or smectic phases 
\cite{Goetz_JCP_98,Marrink_JCB_01,Marrink_JACS_01,Shelley_JCPB_01,Shelley_JCPB_01_2,Lansac_PRE_01,Yoshida_MCLC_01} 
have been limited to smaller system sizes or simulation times,
which were not sufficient to study layer interactions in the smectic phase. 
Schick and coworkers \cite{Netz_PRE_96,Mueller_JCP_96,Mueller_JCP_02}
have investigated one or more polymeric bilayers 
of similar sizes as in the present article, but they focused 
on the formation of pores in the bilayers or fusion between bilayers.
As an alternative, lattice simulations have proven very successful to reproduce
amphiphilic phases \cite{Larson_JP_96,Ciach_JCP_90,Nelson_JCP_97}.
Compared to these, our model avoids lattice artifacts and permits 
to control more easily the surface tension of the bilayers. 

Smectic phases can also be studied with standard coarse-grained models 
for liquid crystals, such as spherocylinders or Gay-Berne particles.
These systems exhibit smectic A phases, which are similar to our
lamellar phase. Otherwise, the phase diagram is rather different.
Liquid crystal models often display a smectic/nematic phase transition, 
which is not present in our model (nor in real amphiphilic systems).
Instead, our model exhibits an anisotropic sponge phase,
and a micellar phase at low amphiphile concentrations.
Nevertheless, we expect that our main results on thermal 
fluctuations in the lamellar phase are also valid for general smectic phases.

\subsection{Simulation details.}

We have studied the model in the $(N P_n P_t T)$-ensemble (constant number of 
particles, constant pressure normal and tangential to the bilayers, and constant temperature) with molecular 
dynamics simulations. 

($N$): The lamellar phase was studied at an amphiphile fraction of $80 \%$ of 
the beads (one solvent bead per $h_2t_2$).
Two system sizes were compared in order to detect finite size 
effects. The small system contained $10\,240$ tetramers and $10\,240$ solvent 
beads, which formed five bilayers of about two thousand molecules 
each. The bigger system was three times larger in the direction of the 
director and contained fifteen bilayers. 
The thermal-averaged box dimensions were $ L_x =  L_y = 43.4 \pm 0.1 \,\sigma$,
 $ L_z = 95.7 \pm 0.2 \,\sigma$ for the system of fifteen bilayers
 and $L_z =  31.9 \pm 0.1 \,\sigma$ for the system of five bilayers.

\begin{figure}[t]
\begin{center}
\includegraphics[width=5.5cm, angle=0]{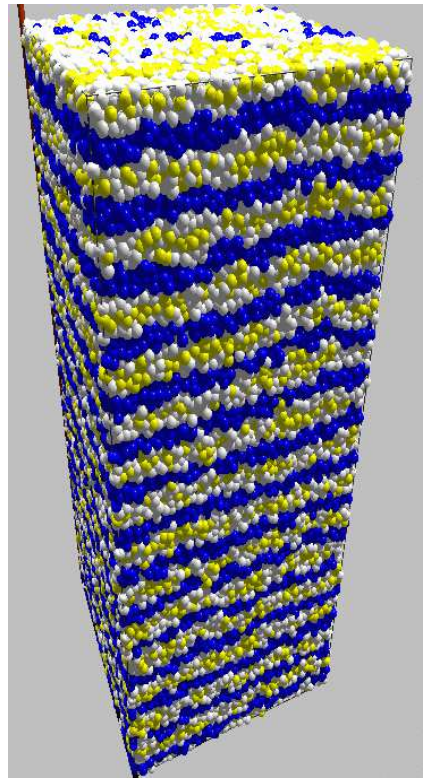}
\caption{Snapshot of a conformation of $30\,720$ $h_2t_2$ tetramers 
and $30\,720$ solvent beads,  simulated in the $NP_nP_tT$ ensemble.
($P^*=2.9$, $T^*=1.0$, $\phi^*=1.1$). The dark beads are
solvophobic (type t), the light beads are solvophilic beads or solvent beads (type h or s). }
\label{m3AS_E11_P29}
\end{center}
\end{figure}

($P$): The normal and tangential pressure component $P_n$ and $P_t$ were
 kept constant using the extended Hamiltonian method 
of Andersen \cite{Andersen_JCP_80,Parrinello_PRL_80}.
The box  shape is constraint to remain a  rectangular parallelepiped. 
The box dimension perpendicular to the bilayer ($L_z$) 
and tangential to the bilayers ($L_x, L_y$) are coupled to two separated pistons.
We imposed separately the two pressure components rather
than the total pressure $P = (P_n+2P_t)/3$ because of technical reasons: 
the mechanical equilibrium  is reached earlier,
the orientation of the bilayers is stabilized, and the surface tension is controlled.
Since we studied a bulk  lamellar phase, 
we imposed an isotropic pressure  ($P_n = P_t= P$). 
More details on the simulation algorithm and simulation 
parameters are given in the appendix A.  

($T$): 
The temperature was controlled with a stochastic
Langevin thermostat, which has been described earlier and applied
to simple and complex fluids by one of us (with coworkers) 
\cite{Kremer_JCP_90,Soddemann_EPJE_01, Kolb_JCP_99}.
The Langevin thermostat leads to the correct temperature
and to the correct canonical distributions of static observables. Therefore
the  thermal fluctuations appearing at thermal equilibrium 
can be studied using such a thermostat.

We define the dimensionless pressure $P^* = P k_BT /\epsilon$,
the dimensionless temperature $T^* = k_BT/\epsilon$ and the dimensionless
 potential depth $\phi^*=\phi/\epsilon$. We have chosen to fix $P^* = 2.9$ and $T^*= 1.0$. 
For the typical value $\phi^* = 1.1$ the densities vary around $0.85$ beads per unit volume. 

With our typical parameters, lamellae form and order spontaneously (see next section). 
However, this process  requires at least $30\,000\,\tau$  in the $NP_nP_tT$ ensemble. 
In most runs, we have 
therefore imposed the orientation of the lamellae to the initial configurations. 
They were constructed such that five or fifteen bilayers, separated by solvent 
layers with always the same number of solvent particles, 
were stacked in the $z$-direction. These configurations were then 
relaxed for $10\,000\,\tau$. During that time, the interlamellar distance adjusted 
to its equilibrium value, the shape of the flexible box changed accordingly,
but the director remained basically aligned with the $z$-direction. 
Fig. \ref{m3AS_E11_P29} shows a snapshot of a large system 
($30\,720$ tetramers and $30\,720$ solvent beads). 
Data for the fluctuation analysis were then collected over $100\,000 \,\tau$
for both system sizes. We verified that the pressure tensor 
obtained at equilibrium was diagonal and isotropic: for example, 
in the simulation of the large system ($P^* =2.9$, $T^* =1.0 $, $\phi^*= 1.1$), 
the averages of the  non-diagonal components  were
 smaller than the errors of the computation (0.01). 
$\langle {\cal P}_{xy}\rangle  = 0.002$, 
$\langle {\cal P}_{xz}\rangle = 0.002$, 
$\langle {\cal P}_{yz}\rangle = - 0.008$, 
$\langle {\cal P}_{xx}\rangle = 2.896 $; 
$\langle {\cal P}_{yy}\rangle = 2.895 $,
in units of $\epsilon /\sigma^{3} $.
We also verified that the ensemble-averaged surface tension 
$\gamma =\langle L_z (P_n-P_t)\rangle$ 
was negligible ($\gamma = - 0.01 \pm 0.01 \, \epsilon\cdot \sigma^{-2}$
per bilayer in the large system).

\section{Results.}

\label{results}

We shall first establish the relevant parts of the phase diagram 
of the model, and then discuss the layer fluctuations in the smectic phase.
The behavior of a pure amphiphile system without solvent has been
studied earlier by Guo and one of us \cite{Guo_JCP_03_1,Guo_JCP_03_2}.
Here we consider systems with solvent particles (20 \% solvent particles
unless stated otherwise).

\subsection{Smectic ordering.}

\begin{figure}[h!]
\begin{center}
\includegraphics[width=8cm]{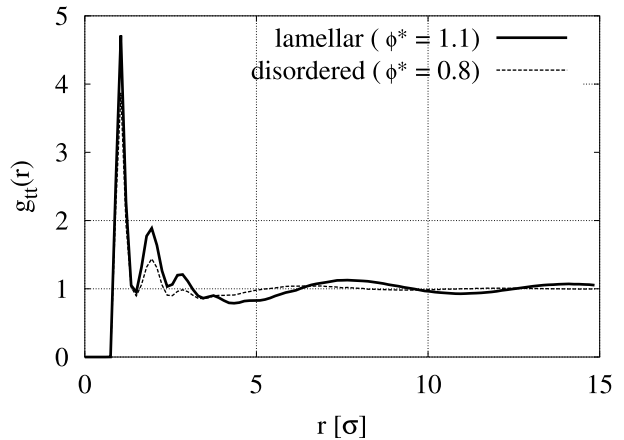}
\caption{Pair correlation function $g_{tt}(r)$ of tail beads ($t$), 
as a function of the distance $r$ in units of $\sigma$measured in a system of 
$10\,240$ amphiphiles $h_2t_2$ and $10\,240$ solvent beads  ($P^*=2.9$,, $T^*=1.0$) .
The two curves correspond to state points in the disordered 
phase ($\phi^*= 0.8 $, dashed line) and in the lamellar 
phase ($\phi^*= 1.1 $, solid line).}
\label{Gr_BB}
\end{center}
\end{figure}

\label{results_phase} 

Fig. \ref{Gr_BB} shows the pair correlation function of tail beads 
$g_{tt}(r)$ in a disordered and in a lamellar state. At short distances, 
it is dominated by the local liquid structure in both cases.
The difference between the two structures becomes apparent at intermediate 
distances: The pair correlation function exhibits small oscillations in
the lamellar phase with a periodicity corresponding to the inter-layer
distance which are not present in the disordered phase. 

The smooth structure of the fluid on intermediate length scales indicates 
the presence of a smectic phase and is usually analyzed in terms of a set
of two order parameters: 
(i) The nematic order parameter, which characterizes the orientational symmetry 
breaking, 
and (ii) a smectic order parameter, which describes the breaking of 
translational symmetry.

\begin{figure}[t]
\begin{center}
\includegraphics[width=8cm]{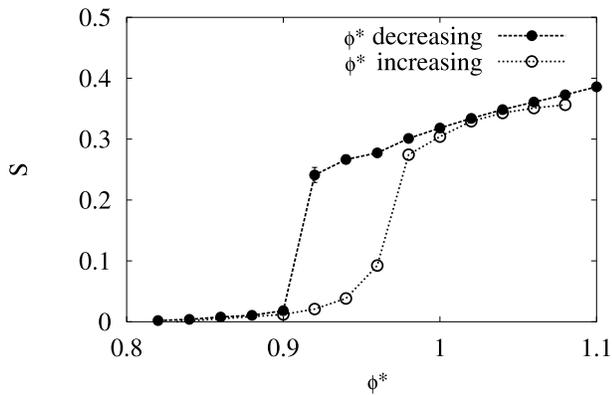}
\caption{Nematic order parameter $S$ as a function of dimensionless depth 
$\phi^*$ ($P^*=2.9$, $T^* =1.0$) in a system of
of $10\,240$ $h_2t_2$ tetramers and $10\,240$ solvent beads. The relative depth
$\phi^*$ was varied from $0.9 $ to $1.1$ and back in steps 
of $0.02$. At each step, the configuration was relaxed over
$5\,000\,\tau$ and the order parameter was then calculated over 
$5\,000\,\tau$. At the ordered side of the transition, the configurations
still contain some linear defects after $10\,000\,\tau$.
Therefore, the order parameter given here is slightly lower than 
the equilibrium value. The lines are guides for the eye. 
}
\label{AS_P29_E_S}
\end{center}
\end{figure}

The nematic order parameter $S$ is the largest eigenvalue of the 
nematic order tensor $\hat{Q}$ 
\begin{equation}
\label{nematic_param}
\hat{Q}_{\alpha,\beta} =\frac{1}{2N} 
\sum_{i=1}^{N}\,\left( 3 u_{i\alpha}u_{i\beta} - \delta_{\alpha \beta}\right)
\text{ with } \alpha,\beta \in \text{ \{x,y,z\} }
\end{equation}
where the ${\bf u}_i$ are unit vectors pointing in the direction of the 
molecules $i$, and the sum runs over all $N$ molecules\cite{deGennes_93}.
 The eigenvector corresponding to the eigenvalue $S$ is the director ${\bf n}$. 
In our analysis, we calculated ${\bf u}_i$ from the direction of the middle
bond of the molecules. Other choices are also possible and lead to similar 
results \cite{Hong-Xia_02}. In particular, the dimensionless potential
depth $\phi^*$ at the order-disorder transition does not depend on 
the details of the analysis. Fig. \ref{AS_P29_E_S} shows 
$S$ as a function of $\phi^*$ for a set of runs where $\phi^*$ 
was increased from 0.82 to 1.1 in steps of 0.02, and then reduced again.
The order parameter jumps from about 0 to $0.38$ when the potential depth 
$\phi^*$ is increased, and back to zero when $\phi^*$ is reduced.
Hysteresis is observed. This indicates the presence of a first order phase 
transition between the disordered phase and an ordered
phase. We did not determine precisely the parameter $\phi^*$ of the transition. Nevertheless,
 Fig. \ref{AS_P29_E_S} clearly shows that the state under investigation ($\phi^*=1.1$)  is in the lamellar phase domain.

\begin{figure}[t]
\begin{center}
\includegraphics[width = 8 cm]{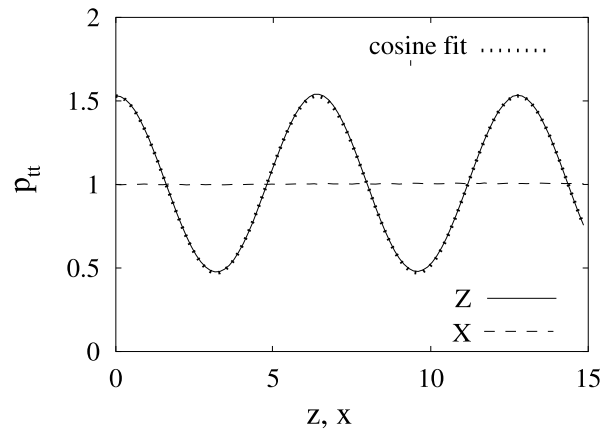}
\caption{Density correlation functions of tail beads in the lamellar
phase ($P^*=2.9$, $T^* =1.0$, $\phi^* = 1.1$)
in the direction $x$ (thin dotted line) and $z$ (thick solid line).
The correlation in $z$-direction is fitted by the function
$f(z) = 1+ \alpha~cos( 2\pi z/\bar{d})$ with  
$\alpha = 0.52$ and $\bar{d} =6.38 \,\sigma$ (thick dashed line).
}
\label{P_BB}
\end{center}
\end{figure}

\begin{figure}[b]
\begin{center}
\includegraphics[width=8cm]{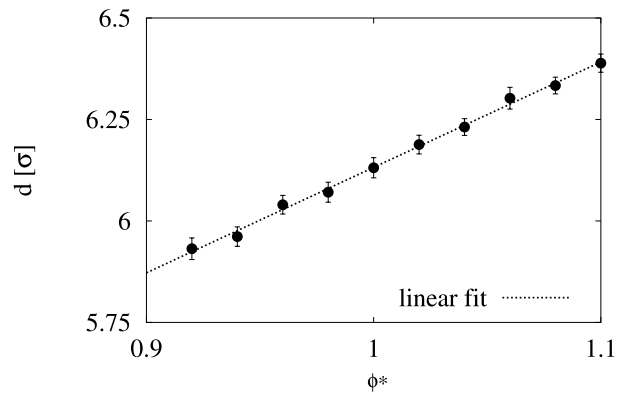}
\caption{Inter-layer distance $\bar{d}$ in units of $\sigma$ as a function 
of potential depth $\phi^*$, calculated from the 
simulation runs with decreasing $\phi$ of Fig. \ref{AS_P29_E_S}.
The line is a linear fit. }
\label{AS_P29_EDOWN_dlam}
\end{center}
\end{figure}

\begin{figure}[t]
\begin{center}
\includegraphics[width=8cm]{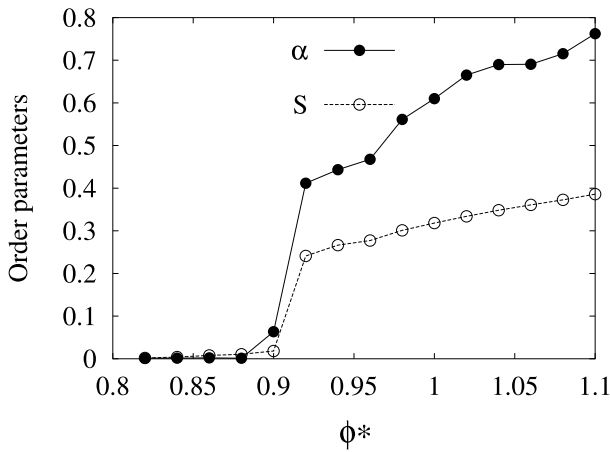}
\caption{Positional order parameter $\alpha$ and orientational order 
parameter $S$ as a function of potential depth $\phi^*$. The data
correspond to the simulation runs with decreasing $\phi^*$
of Fig. \ref{AS_P29_E_S}. 
}
\label{AS_P29_E_alpha}
\end{center}
\end{figure}

The translational symmetry breaking can be investigated by inspection of 
the density-density correlations along the director ${\bf n}$. We divide
the system into slabs of thickness $\Delta z$ in the direction $z$ of the
director, and calculate the density correlations of solvophobic particles 
(t-beads) using
\begin{eqnarray}
 p_{tt}(z) &=&\frac{1}{N_t (N_t -1)} 
\nonumber \\ &&
\times  \sum_{i \neq j}  
\frac{1}{\Delta z} \int_{-\frac{\Delta z}{2}}^{\frac{\Delta z}{2}} \text{d}z'  
\delta\left(\frac{  |z_i-z_j| - [z+z'] }{L_z} \right).
\label{Ptt}
\end{eqnarray}
Here $N_t$ is the number of $t$-beads, $z_i$ and $z_j$ are the 
$z$-coordinates of the beads $i$ and $j$, $L_z$ is the box dimension
in the $z$-direction, and $\delta$ denotes the delta function.
Fig. \ref{P_BB} shows the resulting density correlation function 
for a lamellar state point in the directions parallel and perpendicular 
to the director. The translational symmetry is clearly broken 
in the direction of the director ($z$), and preserved in the other two ($x,y$).
The density oscillations happen to be fitted nicely by a cosine function, 
$f(z) = 1+ \alpha~cos( 2\pi z/\bar{d})$. The period $\bar{d}$ corresponds to
the mean inter-layer distance. As shown in Fig. \ref{AS_P29_EDOWN_dlam}, 
it increases almost linearly with the segregation factor $\phi^*$.
The amplitude $\alpha$ is the order parameter of the translational order. 
It is shown as a function of $\phi^*$ and compared with the nematic
order parameter $S$ in Fig. \ref{AS_P29_E_alpha}. Both order parameters 
are non-zero in the ordered phase, and jump to zero simultaneously at 
$\phi^*\sim 0.92$. As expected for amphiphilic systems
\cite{Morse_PRE_93, Soddemann_EPJE_01},
our model does not exhibit a separate nematic phase.

We have investigated partially the region of stability of the lamellar phase. 
At the number density of about $0.85$ beads per unit volume (which is
the density of a monomeric fluid at the dimensionless pressure $P^*= 3.0$),
a pure system of amphiphiles orders into a lamellar phase at
$\phi^* = 0.77 \pm 0.1$. In contrast, a system which
contains 20 \% solvent beads remains lamellar only down to
$\phi^* \sim 0.98$ (see Fig. \ref{AS_P29_E_S}). The solvent
destabilizes the lamellar phase. At fixed $\phi^*=1.1$,
the maximum amount of solvent beads that could be added to the 
lamellar stack without destroying the smectic order was found to be
roughly 40 \%.  

\begin{figure}[b]
\begin{center}
\includegraphics[width=6cm,angle = -90]{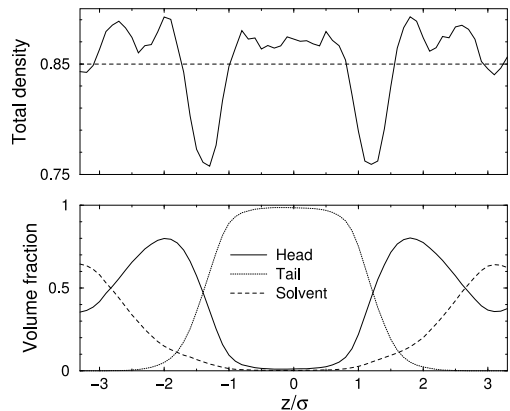}
\caption{Total density profiles (top) and partial volume
fractions of head, tail and solvent beads (bottom) across
a smectic layer ($P^*=2.9$, $T^*=1.0$, $\phi^* = 1.1$).
See text for more explanation.
}
\label{den_profile}
\end{center}
\end{figure}

The main simulations were then carried out at a solvent volume fraction
of 20 \% and $\phi^* =1.1$, which is well in the smectic phase. 
The bead density $\rho = 0.85 \,\sigma^{-3}$ was maintained  
by applying the pressure $P^*=2.9$. 
The local structure of the smectic layers can be characterized by the 
profiles of head, tail, and solvent bead densities. Unfortunately,
the calculation of density profiles is hampered by the fact that
the local positions of the membranes fluctuate both in time and space.
To account for these effects, we have calculated the local positions 
of each membrane on a grid in the $(x,y)$ plane of mesh size 
$1.3\, \sigma$, and evaluated local profiles in the $z$-direction
relative to those positions, which were then averaged.
The procedure is described in more detail in the next section and in 
appendix B. The resulting density profiles are shown in 
Fig. \ref{den_profile}. The solvophobic and the solvophilic beads 
are well segregated. In particular, almost no solvent particles penetrate 
into the amphiphilic bilayers.

\subsection{Fluctuation analysis.}

\label{results_fluc}

\begin{figure}[b]
\begin{center}
\includegraphics[width=6cm,angle=-90]{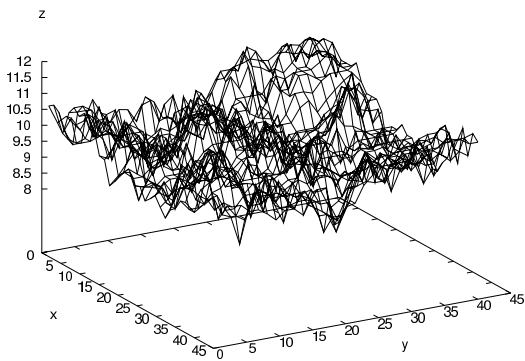}
\end{center}
\caption{Typical conformation of the position $h_n(x,y)$ of a membrane in a 
stack of five membranes. See text and appendix B for explanation.}
\label{lam_pos}
\end{figure}

We now turn to investigate the layer fluctuations in the
lamellar phase. For the fluctuation analysis, we have determined
the local positions of the membranes in every configuration from 
the local densities of solvophobic beads. A volume element is 
considered to be part of a membrane, if the local density of solvophobic 
beads there exceeds a certain pre-defined threshold (between 0.65 and 0.75). 
We characterize the $n$th membrane by its position $h_n(x,y)$ 
and its thickness $t_n(x,y)$ (Monge representation \cite{Safran_94}). 
In practice, only discrete values of $x$ and $y$ were considered 
($x = n_x {L_x/N_x}$ and $y = n_y {L_y/N_y}$). For each point 
$(x,y)$, the position and thickness of a membrane were determined 
as the mean and the difference of the two $z$ values where the local 
density of solvophobic beads crosses the threshold value. The algorithm 
is described in appendix B. The displacement of the layer was then 
defined as $u_n(x,y) = h_n(x,y)-\bar{h}_n$, where the mean position 
$\bar{h}_n$ was determined separately in each configuration 
($\bar{h}_n =\sum_{x,y}  h_n(x,y)  /(N_xN_y)$). 
The two dimensional Fourier transform of $u_n(x,y)$ gives the
fluctuation spectrum (cf. Eq. (\ref{TF2})).
Fig. \ref{lam_pos} shows a typical membrane configuration. 

\begin{figure}[t]
\begin{center}
\includegraphics[width=5.5 cm, angle = -90]{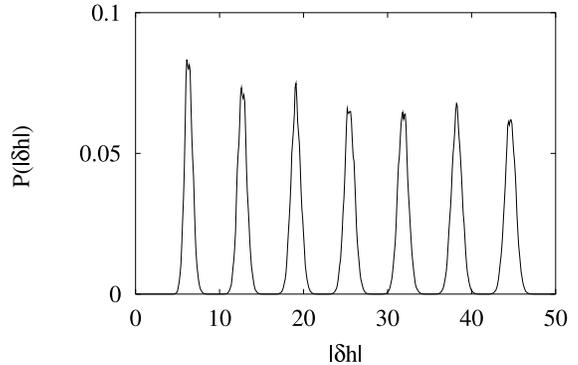}\\
\caption{Distribution of distances between layers in a lamellar stack
of 15 layers.}
\label{M3_P_dh}
\end{center}
\end{figure}

\begin{figure}[b]
\begin{center}
\includegraphics[width=5.5 cm, angle = -90]{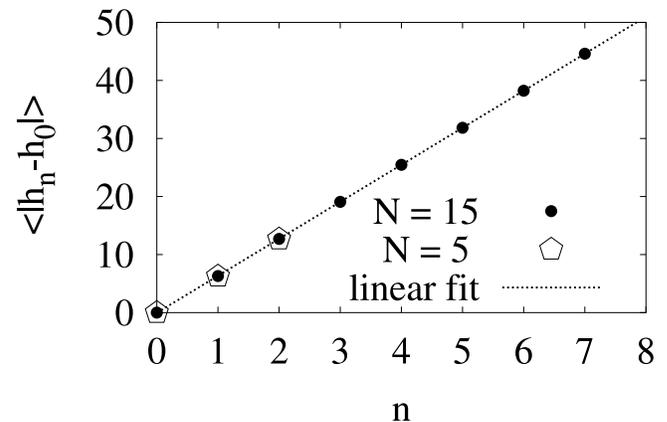}
\caption{Mean inter-lamellar distance between bilayers separated by $n$ 
solvent layers, $\langle |h_n(0)-h_0(0)| \rangle$, vs. $n$. 
The solid line is a linear fit.}
\label{M3_P_dh_mean}
\end{center}
\end{figure}

\begin{figure}[t]
\begin{center}
\includegraphics[width =5 cm, angle=-90 ]{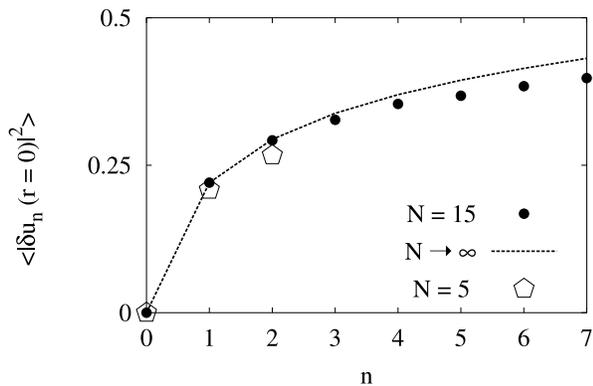}
\caption{Variance of the distribution of inter-lamellar distances between 
bilayers separated by $n$ layers of solvent. Data are shown for the
large system (15 bilayers) and the small system (5 bilayers).
The line is the prediction of Eq. (\ref{delta_unr}), using the
Caill\'e parameter $\eta_1/q_1^2 = 0.055$.}
\label{M3_P_dh_var}
\end{center}
\end{figure}

\label{spatial_fluctuation_sim}
First, we analyze the distribution of inter-layer distances 
$\sum_n \langle |h_n(x,y) - h_0(x,y)| \rangle$. The histogram for
the system of 15 layers is shown in Fig. \ref{M3_P_dh}.
The periodic arrangement of the peaks reflects the smectic order 
of the membranes along the director - the $n$th peak corresponds to the 
distance between bilayers which are separated by $n$ layer(s) of solvent.
For each peak, the mean and the variance were determined by fitting 
a Gaussian function. The results are plotted as a function of $n$ in 
Figs. \ref{M3_P_dh_mean} and \ref{M3_P_dh_var}. Not surprisingly, 
the mean distance is proportional to $n$,
$\langle |h_n(x,y) - h_0(x,y)| \rangle = n \bar{d}$ with 
$\bar{d} =6.38\,\sigma$. The variances reflect the height-height 
fluctuations (cf. Eq. (\ref{delta_unr})). From the width of
the first peak, we calculated the value of the Caill\'e parameter,
$\eta_1 = 0.053$. The variances of the higher order peaks are
compared with the prediction of the continuous theory in
Fig. \ref{M3_P_dh_var}. The theory describes the simulation data 
very well for small $n$. At large $n$, small discrepancies are
observed, which are presumably due to finite size effects:
In infinite systems, $\langle \delta u_n(0)^2 \rangle$ should increase
monotonously with $n$. In a finite system with periodic boundary
conditions, however, it is bound to decrease beyond $n = N/2$
and reaches zero for $n=N$.

Figs. \ref{M3_P_dh_mean} and  \ref{M3_P_dh_var} also display results 
for systems with 5 layers. The interlamellar distance does not depend 
significantly on the system size.  
The variances are reduced in the small system, due to the finite
size effect discussed above. As a consequence, the Caill\'e parameter 
$\eta_1$ is slightly underestimated ($\eta_1 = 0.051$), though still 
of the correct order of magnitude.

\begin{figure}[b]
\begin{center}
\includegraphics[width=6cm,angle=-90]{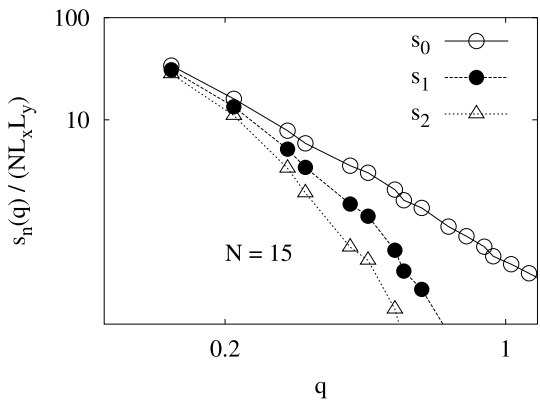}
\caption{Membrane correlation spectra $s_n(q_\perp)/(L_xL_y )$
  vs. $q_{\perp}$ for the system of 15 bilayers.
}
\label{M3_16_corr012}
\end{center}
\end{figure}

\label{fluc_spectra_sim}
Next we study fluctuations and correlations of membranes in the 
$(x,y)$ direction, which we characterize by the quantities 
$s_n({\bf q}_\perp)$ defined in Eq. (\ref{sum_coupled_spectra})
($q_\perp^2 = q_x^2+q_y^2$).  
Fig. \ref{M3_16_corr012} shows  $s_n({\bf q}_\perp)/(L_x L_y)$ 
for $n=0,1,2$ as a function of $q_\perp$. 

Unfortunately, a meaningful comparison of the continuum theory and
the simulation data was only possible for intermediate wavevectors 
$q_\perp$. Large wavelength fluctuations could not be analyzed
reliably because the autocorrelation time for these modes 
($q_\perp = 0.1 \,\sigma^{-1}$) became comparable to the total length 
of the simulation run ($100\,000\,\tau$), even in the small system 
($N=5$ bilayers). The correlation time drops to $2\,500$ $\tau$ for 
$q_\perp = 0.3\, \sigma^{-1}$. On the short wavelength side of the spectrum, 
the continuum theory (\ref{Fd}) breaks down on molecular length scales.
Beyond $q_\perp \geq 1\, \sigma^{-1}$, the fluctuations of the membrane 
thickness have been found to follow a $1/q_\perp^{-2}$ behavior 
\cite{Marrink_JCB_01}. This has been interpreted in terms of an 
effective surface tension caused by the protrusion of molecules out 
of the bilayer. In our system, the protrusion regime is found at 
$q\simeq 0.8 \,\sigma^{-1}$ (data not shown), corresponding to a length scale 
of about $8\,\sigma$. 

For these reasons, we shall restrict our data analysis to the 
$q_{\perp}$ regime $0.3 \,\sigma^{-1} \leq q_\perp \leq 0.8\, \sigma ^{-1}$
in the following. 

\begin{figure}[t]
\begin{center}
\includegraphics[width=5.5cm,angle =-90]{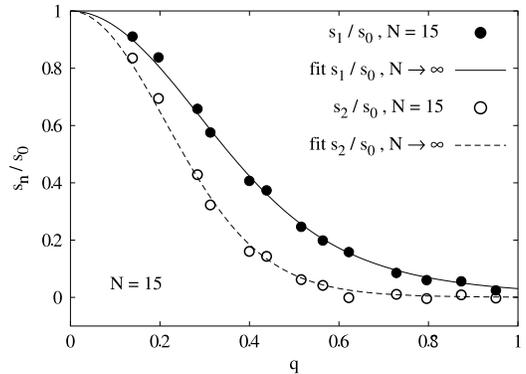}
\caption{Ratio $s_{1,2}(q_\perp)/s_0(q_\perp)$  vs. wave vector $q_\perp$ 
 in the system with 15 lamellae. The dots represent simulation data,
 and the solid lines are fits of Eq. (\ref{sum_coupled_spectra2}) 
 with $X = (\xi q_\perp)^4$ and $\xi_1 = 2.35\,\sigma$, 
 $\xi_2 = 2.33\,\sigma$.}
\label{N15_16_corr012}
\end{center}
\end{figure}

\begin{figure}[b]
\begin{center}
\includegraphics[width=5cm,angle=-90]{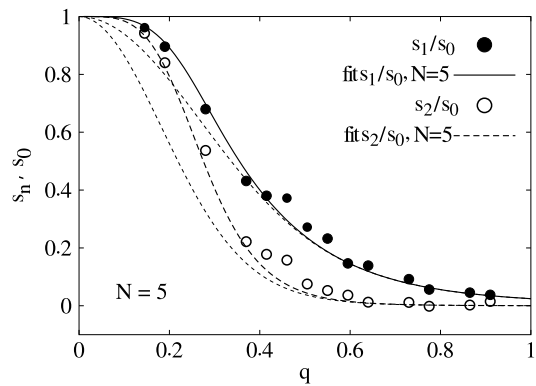}
\caption{Ratio $s_{1,2}(q_\perp)/s_0(q_\perp)$ vs. wave vector $q_\perp$ 
 in the system with 5 lamellae. The dots represent simulation data,
 and the solid lines are fits using  the discrete summation of Eq. 
 (\ref{sum_coupled_spectra1}) with $\xi_1 = 2.5\,\sigma$, $\xi_2= 2.7\,\sigma$.
 Also shown for comparison are the curves obtained with the infinite slab
 approximation (Eq. (\ref{sum_coupled_spectra2}) and the same values
 of $\xi$ (thin lines).
 }
\label{N5_16_corr012}
\end{center}
\end{figure}

\label{cross-correl_sim}
The direct fit of Eq. (\ref{intUq1}) to the data for $s_0(q_\perp)$ in 
this regime was not very significant. Comparing the ratios $s_1/s_0$ and 
$s_2/s_0$ to the theoretical prediction (\ref{sum_coupled_spectra2}) 
turned out to be much more rewarding. In the big system (fifteen bilayers), 
the agreement between our data and the theory is very good 
(see Fig. \ref{N15_16_corr012}). We have fitted the results for 
$s_1/s_0$ and $s_2/s_0$ independently, with only one fit parameter $\xi$. 
Both fits give the same values within the errors, $\xi = 2.34 \pm 0.01\,\sigma$. 
In the small system (five bilayers), the infinite slab approximation 
$N \to \infty$ becomes questionable, therefore we have compared 
$s_{1,2}/s_0$ to the discrete sum obtained by the direct evaluation of 
(\ref{sum_coupled_spectra1}) and (\ref{Uq2}), taking into account
the periodic boundary conditions. At first sight, the agreement seems 
reasonable (see Fig. \ref{N5_16_corr012}). However, the in-plane 
correlation length $\xi$ obtained in the fit, $\xi = 2.6 \pm 0.1\,\sigma$, 
is significantly larger than that calculated in the big system. 
Hence one of the phenomenological parameters $K_c$ or $B$, or both, 
are affected by finite size effects. For example, the effective layer 
compressibility $B$ could be reduced in small systems, due to the fact 
that the layer fluctuations are correlated more strongly 
(cf. Fig. \ref{M3_P_dh}). This would lead to an effective increase 
of the in-plane correlation length $\xi$.
Obviously, the finite thickness of the simulated system
in the direction of the director affects the fluctuations seriously.
In our model, however, a slab thickness of fifteen bilayers seems 
sufficient to recover the behavior described by DH theory for an infinite slab.

\label{auto-correl_sim}

\begin{figure}[t]
\begin{center}
\vspace*{1cm}
\includegraphics[width=6cm,angle=-90]{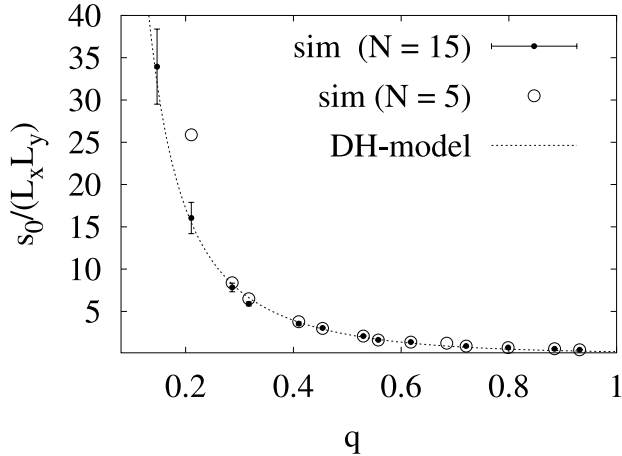}
\caption{Autocorrelation spectrum $s_0(q_\perp)/(L_x L_y)$ of
  membranes in a stack of 15 layers. The solid line 
  gives the prediction of the elastic theory (\ref{intUq1})
  with the parameters $B= 0.13\,k_BT\sigma^{-4}$ and $\xi = 2.35\,\sigma$.
}
\label{S0q}
\end{center}
\end{figure}

From the parameters $\eta_1 = 0.053$ and $\xi = 2.34 \,\sigma$,
we can calculate the bending energy $K_c = 4\, k_BT $ and the compressibility
modulus $B= 0.13 \,k_BT\cdot \sigma^{-4} $. Using these elastic constants and the interlamellar
distance $\bar{d} = 6.38 \,\sigma$, we can now re-inspect the
spectrum $s_0(q_\perp)$ of correlations within single membranes. 
It can be compared directly with the theoretical prediction (\ref{intUq1}), 
without further fit parameter. The result is shown in Fig. \ref{S0q}. 
The discrete harmonic theory describes the data well for the large system. 

\section{Discussion.}
\label{conclusions}

To summarize, we have investigated a bulk lamellar phase in an amphiphilic
system by molecular dynamics simulations, using a phenomenological
off-lattice model of a binary amphiphile-solvent mixture. The system was 
studied in the ($NP_n P_tT$)-ensemble using an extended Hamiltonian, which ensured 
that the pressure in the system was isotropic. Therefore, the membranes 
had no surface tension.

At high amphiphile concentration, (80\% bead percent of amphiphiles), 
the amphiphilic molecules self-assemble into a lamellar phase, {\it i.e.},  
a stack of bilayers. The distances between the membranes fluctuate in a 
way that agrees well with the predictions of the discrete harmonic model. 
From these we could estimate the compressibility modulus $B$ of the smectic.
Furthermore, we have analyzed the in-plane fluctuation spectra of the membranes
and extracted the in-plane correlation length $\xi$. Our results were
in overall good agreement with the predictions of the discrete harmonic
theory, down to wavelengths of roughly $8\,\sigma$.

Fluctuation spectra of free membranes are usually characterized by a scaling
law $s_0(q) \propto q_{\perp}^4$. Looking at the data in Fig. \ref{S0q}, 
we notice that our system does not exhibit a regime with this scaling.
This finding, which was unexpected at first sight, can be rationalized from 
an analysis of the relevant length scales of the system. The elastic theory
(\ref{Fd}) predicts free membrane behavior at wave-vectors $q_{\perp} \xi \gg 1$.
The membrane fluctuations are then incoherent and dominated by 
in-plane correlations.

In our system, however, the validity of the continuum approximation (\ref{Fd}) 
breaks down at wave-vectors larger than $q_\perp \sim \sigma^{-1} \sim \xi^{-1}$, 
and the free membrane regime is never observed. The problem lies in the fact 
that the in-plane correlation length $\xi$ is of molecular order 
($\xi \sim 2.34 \,\sigma$). We recall that $\xi$ is closely related to
the interactions between membranes, which are characterized by the compressibility 
modulus $B$ ($\xi = (K_c/B)^{1/4}$). For free, noninteracting membranes, the
correlation length $\xi$ is infinite. For confined membranes, $\xi$ becomes
finite. In our case, where the lamellae are separated by only a few 
molecules, $(\bar{d} \sim 6 \,\sigma)$, it is not surprising that $\xi$ is
also of the order of the size of molecules. This explains why incoherent
membrane fluctuations cannot be observed in our model. 

We can compare our results to those obtained for a lamellar phase
which is only stabilized by the Helfrich interactions.
Inserting the bending energy $K_c  = 4\,k_BT$, and the membrane
thickness $\bar{t} = 4.4\,\sigma$, one calculates \cite{Roux_JPF_88} 
\begin{equation}
\label{Helfrich_B}
\frac{B}{k_B T} = \frac{9 \pi^2}{64} \frac{(k_BT)}{K_c (\bar{d}- \bar{t})^4} = 
 0.01 \,\sigma^{-4} 
\end{equation}
\begin{equation}
\label{Helfrich_eta1}
\eta_1 = \frac{4}{3} \left( 1-\frac{\bar{t}}{\bar{d}}\right)^2 = 0.08
\end{equation}
We recall that the real compressibility modulus in our model was given by
$B/k_B T = 0.13\, \sigma^{-4}$,  and the real Caill\'e parameter by $\eta_1 = 0.053$. 
Thus the Helfrich theory underestimates the stiffness of the interactions 
between the membranes of our model by one order of magnitude. 

\section*{Acknowledgements}

We thank Thomas Soddeman, Ralf Everaers and Hong-Xia Guo for 
stimulating discussions. We thank the R\'egion Rh\^ones-Alpes for subsidizing 
a generous allocation of computer time on the computing center of the 
Commisariat \`a l'Energie Atomique (Grenoble) within the "conseil des 
partenaires".

\section*{Appendix A:  ($N,P_n,P_t,T$) algorithm.}

Our algorithm was adapted from one published earlier by Kolb 
{\it et al.} \cite{Kolb_JCP_99}, which is similar to the 
piston algorithm proposed by Zhang {\it et al.} \cite{Zhang_JCP_95}.
 It allows to simulate the lamellar phase in a constant-$(N P_n P_t T)$ 
ensemble, where  $P_n$ and $P_t$ are the
pressures normal and parallel to the smectic layers. The ability of
controlling both $P_n$ and $P_t$ was crucial for our simulations, because
the properties of smectic structures depend noticeably on the difference 
between $P_n$ and $P_t$ \cite{Marrink_JCB_01}. Since we simulate a part 
of a bulk lamellar domain, we imposed the same pressure  in both directions $P_n=P_t =P$. 

This was done as follows. As described in the main text, the systems were
set up such that the director of the smectic points along the $z$-direction
of the simulation box. During a simulation run, the director fluctuated
only by a few degrees. Thus the normal and tangential pressure were
essentially given by the diagonal components of the pressure tensor,
${\cal P}_n = {\cal P}_{zz}$ and 
${\cal P}_t = ( {\cal P}_{xx}+ {\cal P}_{yy})/2$.
Here the pressure tensor is defined as usual
\begin{equation}
{\cal P}_{\alpha \beta} =\frac{1}{V} \sum_{i} \frac{m_i v_i^2}{2} \delta_{\alpha \beta} -\frac{1}{6V}
\sum_{i \ne j}{\bf f}_{ij}^\alpha . {\bf r}_{ij}^\beta,
\end{equation}
where $\alpha$ and $\beta$ are $x$, $y$ or $z$;  V is the box volume and
 the sum $i$, $j$ runs over all beads 
in the system, ${\bf f}_{ij}$ is the force exerted by the bead $j$ on the bead 
$i$, and ${\bf r}_{ij} = {\bf r}_i -{\bf r}_j$ is the vector separating
the two beads. 

The constant pressure ensemble was realized using the extended ensemble 
method originally suggested by Anderson, Parinello and Rahman 
\cite{Parrinello_PRL_80, Andersen_JCP_80, Kolb_JCP_99}.
The dimensions $L_{\alpha}$ of the box are taken to be additional
degrees of freedom, which contribute to the Hamiltonian with an
extra kinetic energy and a potential term. 
The extended Hamiltonian then reads
\begin{eqnarray}
 {\cal H}^{ext}& = &
  \left\{ 
	\sum_{i,\alpha} \frac{1}{2 m }  \frac{\pi_{i\alpha}^2}{L_\alpha^2}  
	+ \sum_{i,j>i} v_{ij}(|r_{ij}|) 
\right\}
\nonumber \\ &&
 +\left\{ 
	\frac{1}{2 Q} \left(\frac{\Pi_y^2}{2}+ \Pi_z^2\right) 
	+ P \prod_{\alpha} L_\alpha,
\right\}  
\label{hamiltonien}
\end{eqnarray}
where $Q$ and $\Pi_\alpha$ are the mass and the momenta of the box variables, 
and the other variables refer to the beads: $m$ is the mass, $\pi_{i \alpha}$ 
the momentum  of bead $i$ in the direction $\alpha$, $r_{ij}$ the distance 
between beads $i$ and $j$, and $v_{ij}$ the interaction potential. 
The Hamiltonian defines equations of motion for $L_{\alpha}$ and 
$r_{i \alpha}$, thus the dimensions of the simulation box fluctuate
throughout the simulation. This may cause problems in the $(x,y)$-plane,
where the smectic behaves like a liquid and no mechanism prevents 
excessive deformations of the simulation box. Therefore we have imposed
the constraint that the ratio $\lambda = L_x/L_y$ remains constant during
the simulation \cite{Aoki_PRA_92,Dominguez_MP_02}.
The equations of motion derived from this extended Hamiltonian were
translated into a simplectic algorithm with the direct translation 
technique (see {\it e.g.} Ref.~~\onlinecite{Tuckerman_JPhC_00}). 

The constant temperature was realized by means of a Langevin 
thermostat: We introduced friction forces and a random stochastic
force (noise), with relative amplitudes given by the fluctuation
dissipation theorem \cite{Kolb_JCP_99}.

An actual molecular dynamics update of the algorithm includes
the following steps (the notation is as in Ref.~~\onlinecite{Kolb_JCP_99}: 
$\pi_{i\alpha} = m_i v_{i \alpha} L_\alpha$ and 
$s_{i\alpha} =r_{i \alpha}/L_\alpha$.

\begin{enumerate}
\item
$
\displaystyle
 \forall i, \forall \alpha: 
~ {\bf \pi}_{i\alpha}  \rightarrow  {\bf \pi}_{i\alpha} +
L_\alpha \frac{\Delta t}{2} ({\bf F}_i -\frac{\gamma_p}{L_\alpha m_i} {\bf \pi}_{i}
+\:  \sqrt{k_BT\gamma_p \Delta t}\:
 {\bf \eta}_i(t) 
)
$

\item 
$
\displaystyle
 \Pi_z  \rightarrow   \Pi_z  + \frac{\Delta t}{2} \frac{V}{L_z}({\cal P}_{zz} -P)
\\
\Pi_y  \rightarrow   \Pi_y  + \frac{\Delta t}{2} \frac{V}{L_y}
\left[({\cal P}_{yy} + {\cal P}_{xx}) -2P\right] 
$

\item 
$
\displaystyle
L_z  \rightarrow   L_z  + \frac{\Delta t}{2}  \frac{\Pi_z}{Q}\\
L_y  \rightarrow   L_y  + \frac{\Delta t}{2}  \frac{\Pi_y}{2Q}\\
L_x  \rightarrow  \lambda L_y
$

\item 
$
\displaystyle
\forall i, \forall \alpha: ~  
{\bf s}_{i\alpha}  \rightarrow  {\bf s}_{i \alpha}   
+\frac{ \Delta t }{L_\alpha^2 m_i} {\bf \pi}_{i\alpha}
$

\item  Same as 3.
\item  Calculate new forces and new pressure tensor.
\item  Same as 2.
\item  Same as 1.
\end{enumerate}

We have used the following parameters: $Q =0.1 \,m$,
$\Delta t = 0.005 \,\tau$, $\gamma_p = 1.0\, m\cdot \tau^{-1}$, $T=1.0 \,\epsilon/k_B$.  
The time step $\Delta t = 0.005\,\tau$ was small enough that no bonds could 
break during the simulation.

\section*{Appendix B: Spatial spectral analysis}

This appendix describes how we determined the local positions
of membranes in the lamellar stack.

\begin{enumerate}

\item 
The space is divided into $N_xN_yN_z$ cells of size 
$(dx,dy,dz)$ with $N_x=N_y =32$.
For a density of  0.85 particle per volume unit,  
$dx = dy  \simeq 1.3\,\sigma$ and $dz \simeq 1.0\,\sigma$.
The size of cells may vary from one configuration to another because
the dimensions of the  box  dimensions vary.

\item  
The relative density of tail beads in each cell is calculated
as the ratio $\rho_{tail}(x,y,z) = N_{tail}(x,y,z)/ N_{tot}(x,y,z)$
of the number of tail beads $N_{tail}(x,y,z)$ 
and the total number of particles $N_{tot}(x,y,z)$. 

\item 
The membranes are defined as the space where the relative density of 
tail beads is higher than a threshold ($\rho_{tail}(x,y,z) > \rho_0$).
The choice of the threshold depends on the mesh size in $x-$ and $y-$
directions ($dx = {L_x/N_x}$ and $dy = {L_y/N_y}$). 
Typically, we used 0.65 to 0.75 (80 \% of the maximum relative density
of tail beads).

\item 
The cells that belong to membranes are associated into clusters: Two 
membrane cells that share at least one vortex are attributed to the 
same cluster. Each cluster defines a membrane. This algorithm identifies
membranes even if they have holes. At the presence of necks between
adjacent membranes (local fusion), additional steps have to be taken.
But this happened very rarely in our system.

\item 
For each membrane $n$ and each position $(x,y)$, the two heights 
$h^{min}_n(x,y)$ and $h^{max}_n(x,y)$ where the density 
$\rho_{tail}(x,y,z)$ equals the threshold $\rho_0$ are estimated by a 
linear extrapolation. The mean position and the thickness are then
defined by
\begin{eqnarray}
h_n(x,y) & = & \frac{1}{2} \left[  h^{max}_n(x,y) + h^{min}_n(x,y)\right] \nonumber \\
t_n(x,y)  & = &  \frac{1}{2} \left[  h^{max}_n(x,y) - h^{min}_n(x,y)\right]
\label{mean_pos}
\end{eqnarray}
If the membrane happens to have a hole at $(x,y)$, we attribute the
mean position $\bar{h}_n$ to $h_n(x,y)$ ($h_n^{hole}(x,y) = \bar{h}_n$). 
If two neighboring membranes $i$ and $j$ are connected by a neck at 
$(x,y)$ (local fusion), both membrane positions are taken to be at
$h_{i,j}^{neck}(x,y) = \left[  h^{max}_i(x,y) + h^{min}_j(x,y)\right]/2$. 

\item 
The functions $u_n(x,y) = h_n(x,y)- \bar{h}_n$ are calculated and Fourier
transformed in the $x$ and $y$ dimension as defined by the Eq. (\ref{TF2}),
giving $u_n(q_x,q_y)$. The correlation functions $s_n(q_x,q_y)$ are calculated 
via Eq. (\ref{sum_coupled_spectra1}). The radial average of $s_n(q_x,q_y)$, 
$s_n(q_\perp)$, is performed by binning over wave-numbers on a grid which does 
not depend on the dimension of the box. Ensemble averages were carried out 
for $s_n(q_\perp)$ $(n=0,1,2)$.

\end{enumerate}

\newpage

\bibliographystyle{../styles/prsty}
\bibliography{claire1}

\end{document}